\documentclass{aastex62}
\usepackage{natbib}
\usepackage{amsmath}

\graphicspath{{./}{figures/}}

\received{January 1, 2018}
\revised{January 7, 2018}
\accepted{\today}
\submitjournal{ApJL}

\shorttitle{WFIRST and EUCLID: measuring the mass of free-floating planets}
\shortauthors{Bachelet et al.}
\begin{document}

\title{WFIRST and EUCLID: enabling the microlensing parallax measurement from space.}

\correspondingauthor{Etienne Bachelet}
\email{etibachelet@gmail.com}

\author[0000-0002-6578-5078]{Etienne Bachelet}
\affil{Las Cumbres Observatory, 6740 Cortona Drive, Suite 102,93117 Goleta, CA, USA}

\author[0000-0001-7506-5640]{Matthew Penny}
\affil{Department of Astronomy, The Ohio State University, 140 West 18th Avenue, Columbus, OH 43210, USA}



\begin{abstract}
The Wide Field Infrared Survey Telescope (WFIRST) is expected to detect hundreds of free-floating planets, but it will not be able to measure their masses. However, simultaneous microlensing observations by both Euclid and WFIRST spacecraft, separated by ${\sim}100,000$~km in orbits around the Sun-Earth L2 Lagrange point, will enable measurements of microlensing parallax for low-mass lenses such as free-floating planets. Using simple Fisher matrix estimates of the parallax measurement uncertainties, we show that high-cadence observations by Euclid could be used to measure ${\sim}1$ free-floating planet microlens parallax per 6 days of simultaneous Euclid observations. Accounting for Euclid's pointing constraints, it could therefore potentially measure ${\sim}20$ free-floating planet parallaxes with 120~days of observations split equally between Euclid's main mission and an extended mission, with a potential to increase this number if spacecraft pointing constraints can be relaxed after the end of the main mission. These Euclid observations would also provide additional mass measurements or cross-checks for larger numbers of WFIRST's bound planets, among other benefits to several science cases.
%In the near future, the WFIRST and EUCLID missions can reveal an unprecedent amount of microlensing planets and therefore dramatically improve our knowledge of cold planets. But the primary observable in a microlensing binary event is the mass ratio $q$ between the components. To derive the total mass of the system, it is essential to measure the microlensing parallax. However, it has been shown recently that Gaia-like orbits around L2 are not suitable for the constraint of the microlensing parallax, especially for the lighter lenses.
%In this letter, we present the benefits of complementary observations of the WFIRST and EUCLID missions. 

\end{abstract}

\keywords{editorials, notices --- 
miscellaneous --- catalogs --- surveys}


\section{Introduction} \label{sec:intro}

Gravitational microlensing is a formidable tool with which to explore the population of non-luminous objects. Originally applied 
to the detection of Massive Compact Halo Object (MACHO) dark matter \citep{Paczynski1986,Aubourg1993,Alcock1993,Udalski1993}, microlensing surveys have significantly increased their sky coverage and observing cadence in the past decade in order to search for exoplanets \citep{Gaudi2012}.
This has lead to the discovery of a wide variety of bound exoplanets \citep[e.g.,][]{Bond2004,Beaulieu2006}, brown dwarfs \citep[see, e.g.,][and references therein]{Bachelet2019}, binary stars \citep[e.g.][]{Han2018} and stellar remnant candidates \citep[e.g.,][]{Shvartzvald2015,Wyrzykowski2016}.

Microlensing is especially powerful to detect isolated planetary mass object, the so-called free-floating planets \citep{Oasa1999,Zapatero2000,Luhman2005,Burgess2009,Marsh2010}. \citet{Sumi2011} claimed the discovery of a large population of unbound or large separation (i.e. $\ge100$ AU) planets from measurements of the microlensing timescale distribution. This population of Jupiter-mass objects would have to be twice as common as main-sequence stars, which is an occurrence rate about an order of magnitude larger than predicted by planet formation models~\citep{Veras2012,Ma2016}. However, more recent timescale distribution measurements \citep{Mroz2017,Mroz2019} refute a large population of massive planets but indicates a potential population of less massive free-floating planets that would be observable with space-based microlensing surveys~\citep[e.g.,][]{Barclay2017}.

%Changed text:
%Microlensing is especially powerful to detect isolated planetary mass object, the so-called free-floating planets (Oasa et al. 1999; Zapatero Osorio et al. 2000; Luhman et al. 2005; Burgess et al. 2009; Marsh et al. 2010).  Sumi et al. (2011) discover a large population of unbound or large separation (i.e. ≥ 100 AU)  planets  from  microlensing  timescale distribution.   This  population  of  Jupiter-mass  objects  was reported to be twice as common as main-sequence stars.  This is in strong disagreement with planet formation models that predicts an occurrence at least a magnitude smaller (Veras & Raymond 2012; Ma et al. 2016).  The latter is confirmed by more observations (Mr ́oz et al. 2017; OGLE Collaboration et al. 2019), which reject the large population of massive planets but indicates a potential population of telluric free-floating planets observable from space (Johnson et al., in prep.).

In many cases the clear interpretation of the microlensing event has relied on mass-distance relations from the detection of microlensing parallax $\pi_{\rm E}$, and measurement of the angular Einstein ring radius $\theta_{\rm E}$. Together these observables allow the measurement of the distance of the lens $D_{\rm l}$ and the mass of the lens~\citep[e.g.,][]{Gould2000}
\begin{equation}
M_{\rm l} = {{\theta_{\rm E}}\over{\kappa\pi_{\rm E}}},
\end{equation}
where $\kappa = 8.144 ~\rm{mas.M_{\odot}^{-1}}$. 
Without constraints on the microlens parallax or angular Einstein radius, the only information about the mass of the lens is provided by the Einstein radius crossing time, or more succinctly the event timescale
\begin{equation}
t_{\rm E} = \frac{\theta_{\rm E}}{\mu_{\rm rel}},
\end{equation}
from the lightcurve of the microlensing event, where $\mu_{\rm rel}$ is the relative lens-source proper motion. The angular Einstein radius
\begin{equation}
    \theta_{\rm E} = \sqrt{\kappa M \frac{1-D_{\rm l}/D_{\rm s}}{D_{\rm l}}},
\end{equation}
depends on the mass as well as the lens and source distances, $D_{\rm l}$ and $D_{\rm s}$, respectively, so the timescale is a degenerate combination of the lens mass, distances and the relative proper motion of the lens and source. 
%\begin{equation}
%t_E= {{\theta_E}\over{\mu_{rel}}}    
%\end{equation}
%where $t_E$, the Einstein ring crossing time, is generally well constrained by the microlensing light curve.

Often, $\theta_{\rm E}$ is derived from finite source effects visible in the light curve \citep{Witt1994}, but can also be measured with high-resolution imaging~\citep[e.g.][]{Bennett2007}.
Microlensing parallax can be measured by the subtle effect of an accelerating observer on the lightcurve (so-called annual parallax), but generally requires timescales of $\gtrsim 50$~days for the effects to be detectable~\citep[e.g.,][]{Poindexter2005}. Alternatively, observations of a microlensing event from two well separated locations (${\sim}$AU), so-called satellite parallax, can yield a measurement of microlensing parallax~\citep[e.g.,][]{Refsdal1966,Gould1994,Henderson2016,Zhu2017}. The measurement of the microlensing parallax alone is also a precious tool to explore the mass distribution towards the Galactic Bulge \citep{CalchiNovati2015}. Note that accurate photometry is of first importance to obtain strong constraints on the microlensing parallax from space, as recently pointed by \citet{Koshimoto2019}.
 Two of the primary goals of the Wide Field Infrared Survey Telescope (WFIRST) space mission are to study dark energy via wide-field imaging and spectroscopic surveys~\citep{Spergel2015}, and to conduct a large statistical survey of exoplanet demographics using microlensing \citep{Penny2019}. The mass of most of WFIRST bound planets and their hosts will be estimated via measuring the amplitude and direction of the relative proper motion and the measurement of the lens flux with high resolution imaging~\citep[e.g.,][]{Bhattacharya2018}. However, some planet hosts will be too faint, or will not move far enough away from bright source stars, to detect. In these cases, microlens parallax can provide a mass measurement, and even if lens light is detected a parallax measurement can provide a vital cross check on the len-light-derived mass, which can be contaminated by light from other stars. 
 
 For free-floating planets, WFIRST alone will not be able to measure their masses because they will emit essentially no light, and their ${\lesssim}1$~day event timescales are too short to measure annual parallaxes (see below). However, in many of its free-floating planet events WFIRST will detect finite source effects and measure $\theta_{\rm E}$ (Johnson et al., in prep.). Therefore, if their microlens parallax can be measured via satellite parallax, it is possible to measure the free-floating planets' masses. Unfortunately, the short timescale of free-floating planet events makes it impossible to alert follow-up observations in time to catch the event, so only wide-field high-cadence observations can be used for this task. While simultaneous ground-based observations with WFIRST have been suggested \citep[e.g.,][]{Zhu2016,Street2018}, they face extraordinary challenges to be able to effectively observe the extremely faint microlensing events that WFIRST will find (see Figure~\ref{fig:mag_te}), even for the Large Synoptic Survey Telescope \citep{Abell2009} if WFIRST fields are selected at in high-extinction regions.
 
 Euclid, the European Space Agency mission with a singular goal of characterizing dark energy via wide-field surveys of the extragalactic sky~\citep{Laureijs2011}, has similar wide-field imaging capabilities to WFIRST. \citet{Penny2013} showed that Euclid will be well suited to conducting exoplanet microlensing searches, and significantly more capable than ground-based searches. 
 %The benefits of simultaneous observations from multiple platforms is already known, for example to break microlensing parallax degeneracies \citep{Gould2004,Zhu2017}. 
 In this letter, we show how a modest investment of Euclid's time to obtain observations simultaneously with WFIRST's microlensing survey observations is the only way to measure parallaxes of microlensing events due to telluric free floating planets predicted by star formation theories.
 
%In the Section~\ref{sec:notations}, we quickly review the formalism and approximations made. We then present our results and conclude.
%\section{The Fisher matrix formalism and hypothesis} \label{sec:notations}
\section{Fisher matrix parallax uncertainties and observational parameters} \label{sec:notations}

To study the potential of simultaneous Euclid and WFIRST microlensing observations, we used the Fisher matrix formalism \citep{Gould2013,Mogavero2016,Bachelet2018} to estimate the precision of parallax measurements of microlensing events observed by both spacecraft. Based on this formalism, \citet{Mogavero2016} defines the minimum error parallax measurement $\sigma_{\pi_{E,min}}$ as:
\begin{equation}
\sigma ^2_{\pi_{E}, min}(\phi)= {{\sigma_{\pi_\parallel}^2+\sigma_{\pi_\perp}^2}\over{2}} - {{\sqrt{(\sigma_{\pi_\parallel}^2-\sigma_{\pi_\perp}^2)^2+4~\rm{cov}(\pi_\parallel,\pi_\perp)^2}}\over{2}}
\end{equation}
with $\sigma_{\pi_{\parallel}}$, $\sigma_{\pi_{\perp}}$ and $cov(\pi_{\parallel},\pi_{\perp})$ the error on the parallel and perpendicular components of the parallax vector and their mutual covariance. 
To compute the Fisher covariance matrix we assume that the microlensing events are single lens events with no finite source effects.\footnote{The assumption of no finite source effects does not always apply, but every case where it does not (either high magnification or low planet masses) is generally favorable to measuring parallax}. Following \citet{Mogavero2016}, we assume that the photometric noise level $\sigma$ at the baseline is a free parameter. We also assume that there is no blending from isolated stars (but, as described below, we do include a smooth background due to unresolved stars when estimating photometric precision).
%We simulated the noise model of WFIRST and EUCLID in a wide magnitude range.

%Check VIS data rate.

We assume WFIRST observations are taken continuously within an observing window of 70 days. We consider a microlensing events with a peak time $t_0$ set to be the center of the WFIRST observations. Euclid will have more restrictive observing constraints and can only observe the same fields as WFIRST continuously for 20 days in a window centered 10 days earlier or later than WFIRST's window center \citep{GomezAlvarez2018}. %We note that this cadence can create biases in our analysis. In fact, if $h$ is a divisor of the observing windows length, an observing point is always obtained at the peak. To diminish this effect, we deleted the observation closest to the peak of event from our EUCLID simulations. Therefore, our study is conservative.

Both telescopes are assumed to be orbiting the L2-point with a circular orbit of radius $R=300000$ km, an orbital period $P$ of 180 days and separated by a phase $\phi$. These parameters are broadly representative of spacecraft orbits around L2, and we will consider the impact of changing the spacecraft separation in Section~\ref{sec:results}.
We consider the microlensing source to be in the Galactic Bulge (i.e. $D_{\rm s}=8$~kpc) and a relative source-lens transverse velocity $V = 200$ km/s. The angle $\theta$ between the source trajectory and the projected East vector on the plane of sky is set to $45\deg$ (see \citet{Bachelet2018} for more details on the geometry of the problem). The latitude of the microlensing event relative to the orbital plane of the spacraft about L2 is set to $30 \deg$. Finally, the lens is assumed to be at $4$~kpc, leading to a direct relation between the event timescale, the microlens parallax,
\begin{equation}
\pi_E= 4.3 \biggl({{1 ~\rm{d}}\over{t_E}}   \biggr),
\end{equation}
and the lens mass
\begin{equation}
    M = 0.87 M_{\rm Jup} \left(\frac{t_{\rm E}}{1~{\rm d}}\right)^2.
\end{equation}
Note that the lens distance has a significant impact on the magnitude of the parallax; our choice of a single value is designed to simplify the parameter space while providing a reasonable quantitative estimate of parallax detectability.

\citet{Bachelet2018} consider the L2-point to be inertial and ignore the effect of the annual parallax. To challenge this assumption we use equation 16 of \citet{Gould2013} to estimate the uncertainty of a $1$-dimensional annual parallax measurement for a short, reasonably high-magnification microlensing event. We can combine this with our rough scaling for $\pi_{\rm E}$, and the cadence of WFIRST observations, to estimate the minimum event timescale for which annual parallax can provide interesting constraints. \citet{Bachelet2018} also shown than the Fisher matrix formalism can be optimistic, therefore we require a 5 $\sigma$ constraint on the the 1-d parallax, i.e.  $5\sigma_{\pi_{\rm E},{\rm 1-d}}<\pi_{\rm E}$. The annual parallax constraints are only possible if
\begin{equation}
    t_{\rm E} \gtrsim 3.6~\text{d}~\left(\frac{\sigma}{0.01~\text{mag}}\right)^{2/3},
\end{equation}
where $\sigma$ is the photometric precision of WFIRST at the baseline. Therefore, the acceleration of the Lagrangian point L2 around the Sun is insufficient to constrain the parallax for low-mass lenses, especially for free-floating planets. To confirm this, we apply the Fisher formalism to the motion of WFIRST around the Sun. The detection limits of microlensing parallax measurements by WFIRST alone due to annual parallax are shown in red in Figures~\ref{fig:eucl_wfirst} and \ref{fig:mag_te}. These calculations confirm that WFIRST alone will not measure the parallax for events with $t_E\le 4$ days.

\section{WFIRST and EUCLID noise model}\label{sec:simu}

We assume that WFIRST observations will be taken in the $W146$ filter every 15 minutes with an exposure time of 47~s~\citep{Penny2019}. Euclid has two instruments that can simultaneously observe the same field through a dichroic~\citep{Laureijs2011}. The VIS instrument is an optical imager with 0.1" pixels and a wide-bandpass filter roughly covering $r, i$, and $z$ bands~\citep{Cropper2018}; the NISP instrument is a near infrared imager with 0.3" pixels and wide $Y$, $J$, and $H$ filters~\citep{Maciaszek2016}. We consider two observing modes for Euclid: shallow, high-cadence observations, and deep, low-cadence observations. For the high-cadence observations we assume an exposure time of 100~s in both VIS and NISP-$H$ conducted with a cadence of 30 minutes; with Euclid's expected adjacent field slew times of ${\sim}350$~s~\citep{GomezAlvarez2018} this cadence allows four fields to be observed covering most of WFIRST's nominal fields from \citet{Penny2019}.\footnote{This would generate ${\sim}2100$~Gbit$/$day of data before compression, so Euclid's on-board compression algorithms would need to achieve a compression ratio of ${\sim}2.5$ for crowded bulge images in order for the data to fit within Euclid's regular 850 Gbit/day daily downlink budget; for comparison the \texttt{RICE} and \texttt{HCOMP} compression algorithms of \texttt{fpack} achieved compression ratios of 3.0 and 2.9, respectively, on a simulated 16~bit VIS image of the bulge from \citet{Penny2013}.} For the deep observations we assume that a field would be observed at four dither points each with an exposure time of 250~s in each instrument, requiring a total of just under 30 minutes of observations for each field. We will consider the possibility of these deep observations being taken at a range of cadences, which would represent observations used to fill gaps in Euclid's observing schedule.

The signal to noise of WFIRST and Euclid observations were calculated assuming a microlensing source star is drawn from a 10-Gyr, [Fe/H]$=0.0$ MIST isochrone with WFIRST magnitudes~\citep{Dotter2016,Choi2016,Paxton2011} at a distance of 8~kpc. Euclid magnitudes were assumed to be equal to PanSTARRS $i$ for VIS, and 2MASS $H$ for NISP-$H$, with AB and Vega zeropoints handled appropriately. Extinction as a function of position was drawn from the map of \citet{Gonzalez2012}, and converted to other bands using the \citet{Cardelli1989} extinction law with $R_V=2.5$~\citep[e.g.,][]{Udalski2003}. Crowding was accounted for by adding a smooth background light component with a surface brightness set by the combined brightness of all stars fainter than magnitude $m_{\lambda}$ such that there is fewer than one star per photometric aperture brighter than $m_{\lambda}$, with stars drawn from a population synthesis Galactic model (Huston \& Penny in prep.). Parameters for Euclid's and WFIRST's instruments were taken from \citet{Penny2013} and \citet{Penny2019}, respectively, and references therein.

\begin{figure}[h]
\centering
\includegraphics[scale=0.5]{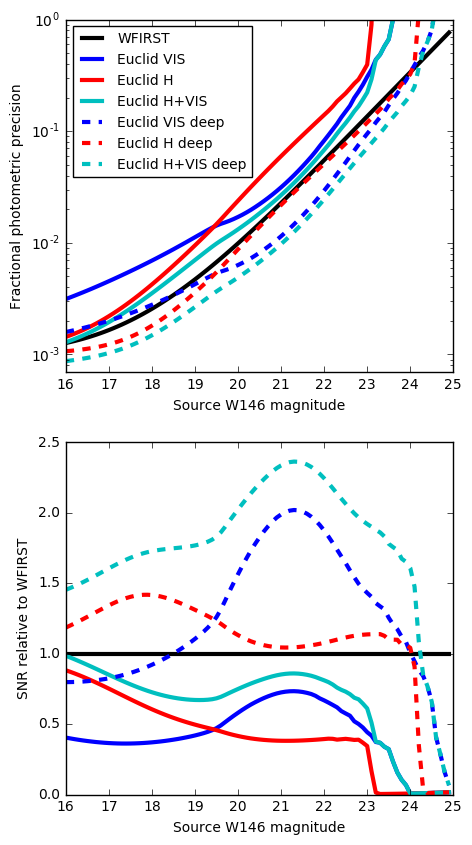}
\caption{Comparison of the photometric precision of Euclid's VIS (blue lines) and NISP-$H$ (red lines) observations and their combined precision (cyan lines) to a standard WFIRST microlensing observation (black lines). The top panel shows the fractional photometric precision as a function of WFIRST $W146$ source magnitude, accounting for intrinsic source color as determined from a bulge star isochrone, extinction, blending, sky and thermal backgrounds, readout noise, and a systematic noise floor. The bottom panel shows the signal to noise ratio of Euclid observations relative to WFIRST. VIS performs better than NISP-$H$ for bluer main sequence stars closer to the sky background, while NISP-$H$ performs better for redder giant stars above the background. The combined Euclid precision is calculated by combining VIS and NISP measurements in quadrature.}\label{fig:noise-model}
\end{figure}

To account for variations in stellar density and extinction, signal to noise curves were computed for a number of line of sight within the nominal WFIRST fields and averaged. To be used in the average, the signal to noise ratio for each instrument needed to be larger than 3. The combined Euclid $H$ and VIS signal to noise was computed by addition in quadrature if both channels produced $S/N>3$ before averaging. The average signal to noise curve for WFIRST and each of Euclid's instruments and considered observing modes is shown in Figure~\ref{fig:noise-model}, as well as the ratio of Euclid's to WFIRST's signal to noise ratios. Although there are variations as a function of magnitude, for the majority of WFIRST's main sequence source stars ($W146{\sim}20$--$24$), Euclid's fast cadence mode would reach a $S/N$ that is a factor of ${\sim}60-85$\% that of WFIRST brighter than $W146<23$, while the deep observations would be a factor of $1.8$--$2.3$ better than a single WFIRST observation. For main sequence bulge stars, Euclid's VIS instrument can produce better photometry than the NISP instrument due to its finer pixel scale and lower backgrounds. For simplicity we assume the combination of VIS and NISP observations results in a Euclid high-cadence precision is a factor of $1.33$ larger (i.e., worse) than WFIRST's, and its deep photometric precision is $0.5\times$ WFIRST's (i.e., better).

\section{Results} \label{sec:results}

\begin{figure}[ht!]
\plotone{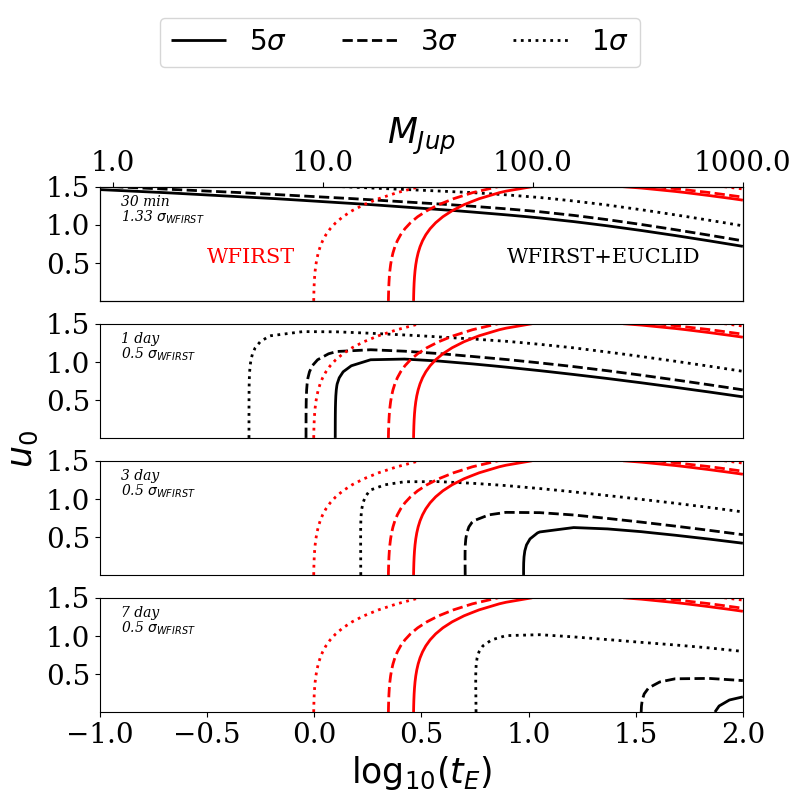}
\caption{The constraints on microlensing parallax provided by simultaneous Euclid and WFIRST observations as a function of event timescale and impact parameter for several possible Euclid observing cadences, for a $W146=20$ mag baseline event (i.e. $\sigma_{WFIRST} = 0.01$. Small-dashed, long-dashed and solid lines indicate the maximum impact parameter for which it is possible to make a parallax detection at 1,3 and 5 $\sigma$ confidence. Black curves represent WFIRST and Euclid simultaneous observations, while red curves represent WFIRST alone. Note that the WFIRST alone represents the detection through annual parallax while the black lines represent the detection through the rotation around L2 only (i.e., annual parallax is ignored). These contours are proportional to the fraction of events for which parallax is measurable. The EUCLID cadence and baseline uncertainty are indicated in the top left hand corner of each plot. \label{fig:eucl_wfirst}}
\end{figure}

Our goal is to test the capability of the simultaneous microlensing observations by Euclid and WFIRST to constrain microlensing parallaxes, relative to the constraints that WFIRST alone would provide. Given a range of possible scheduling constraints, we simulate both high-cadence (30-min), continuous Euclid observations and occasional, deep Euclid measurements with three possible cadences, namely 1, 3 and 7 days. We begin by studying an ideal case where the orbital phase separation between the Euclid and WFIRST is maximal, i.e. $\phi \sim \pi$. We assume a 20th magnitude source with photometric precision of $0.01$~mag in WFIRST and $0.01$~mag in combined Euclid VIS and NISP photometry. Figure~\ref{fig:eucl_wfirst} shows contours of the fractional parallax uncertainty as a function of event impact parameter and event timescale. Contours in the plot show the locations of $1$, $3$, and $5$-$\sigma$ parallax measurements, such that events with impact parameters below the contour have significant parallax measurements. Red contours show the significance of the parallax measurement for only WFIRST observations, while black contours show the significance for both Euclid and WFIRST measurements. At a given timescale, the $y$-axis contour position is proportional to the fraction of events for which parallax measurements are possible, i.e., if one only considers events with $u_0<1$ then the contour is the fraction of events for which parallax is measurable. The plotted results show that low-cadence Euclid observations taken at least as frequently as one observation per day can enable the measurement of microlensing parallax for some events with stellar-mass lenses and timescales $<4$~days, which is an improvement over WFIRST alone. However, only ${\sim}15$\% of microlensing events are bright enough and only ${\sim}10$\% have small enough impact parameters for these constraints, but those with the brightest sources are least likely to have direct lens detection masses because the source is more likely to outshine the lens.

\begin{figure}[h]
\centering
\includegraphics[width=0.5\textwidth]{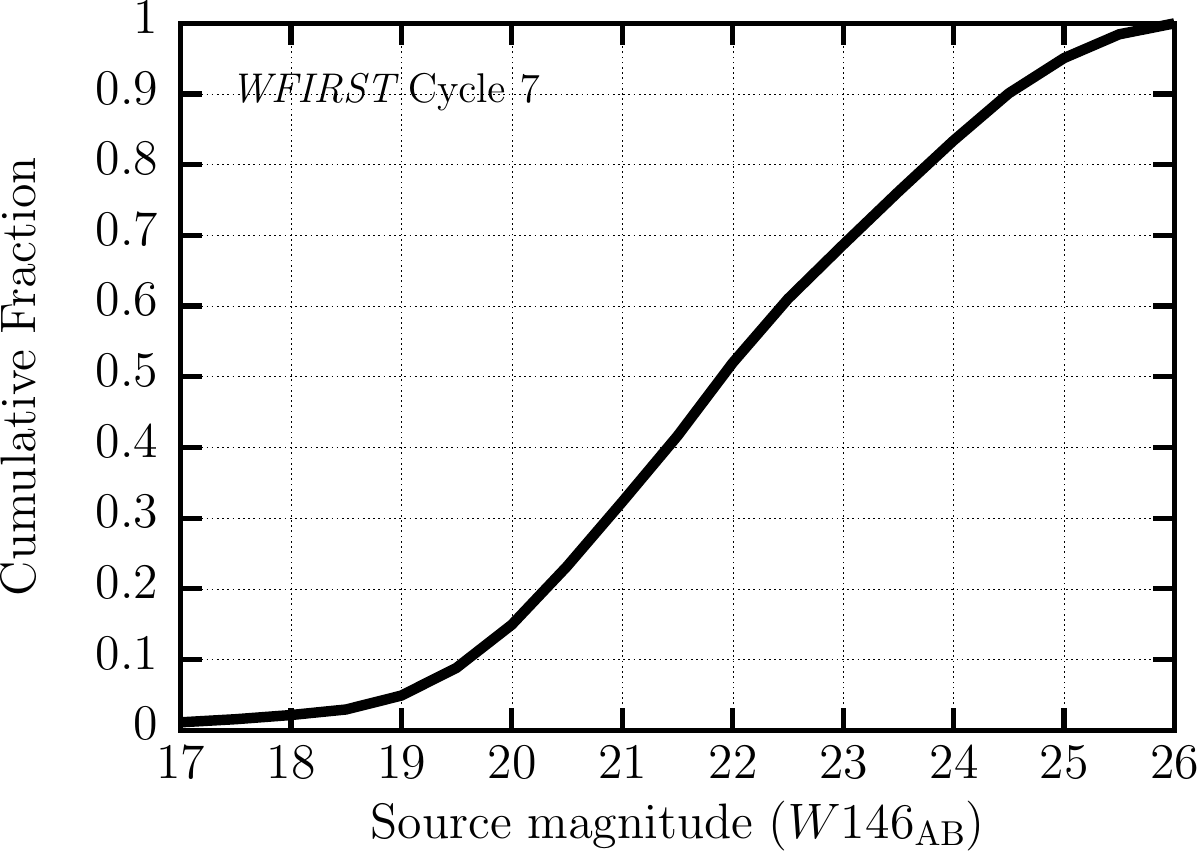}
\includegraphics[width=0.5\textwidth]{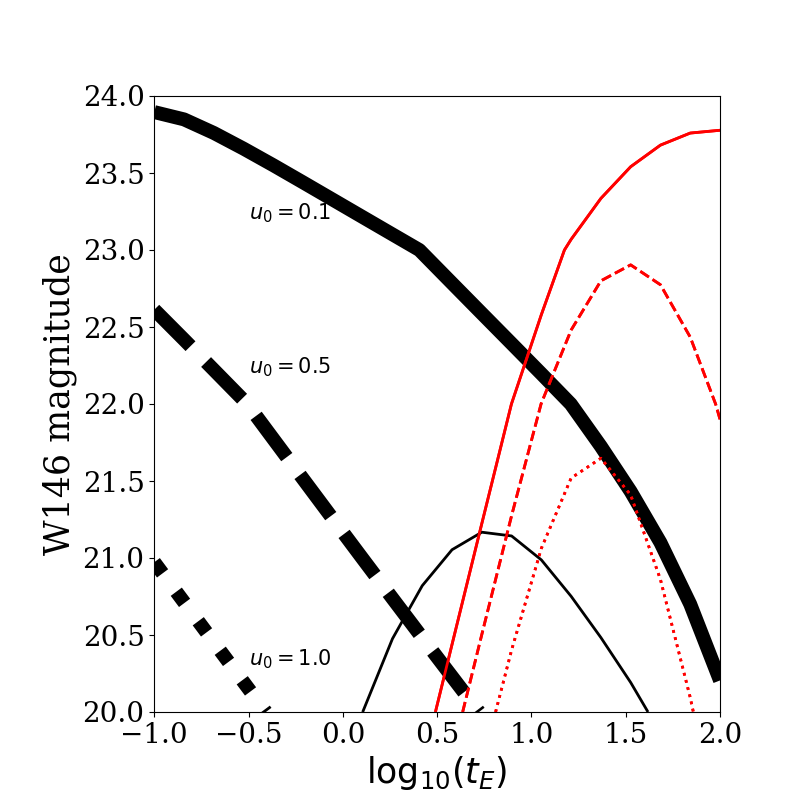}
\caption{{\it Top}:  Cumulative distribution of microlensing source magnitudes from \citet{Penny2019} for events with planet detections. {\it Bottom}: Microlensing parallax detection limits as a function of the event timescale $t_E$ and the source magnitude $W146$. Black lines represents the 5 $\sigma$ detection zones for the WFIRST and Euclid simultaneous observations. Thick lines represent 30 min cadence for Euclid, the thin line is 1 day cadence. Plain, long-dashed and small-dashed lines are for $u_0$ equals to 0.1, 0.5 and 1.0 respectively. The red lines represent the 5 $\sigma$ detection zones of WFIRST alone through annual parallax.\label{fig:mag_te}}
\end{figure}

Beyond demonstrating detectability, it is valuable to estimate the number of events for which Euclid can provide parallax constraints. The top panel of the Figure~\ref{fig:mag_te} shows the cumulative distribution of WFIRST source magnitudes for events with planet detections from \citet{Penny2019}; the majority have source magnitudes in the range $W146\approx20$--$24$. The bottom panel shows a map of parallax precision plotted as a function of source magnitude and event timescale assuming several impact parameters and shallow Euclid observations with a cadence of 30  min or deep observations with a cadence of 1 day. The deep 1-day cadence observations would provide parallax constraints for fewer than the brightest 30\% of events with $u_0<0.1$ (i.e., of the order of a few percent of events), and it is unlikely that Euclid would have significant holes in its schedule every day, so we do not consider deep, low-cadence observations further.

In contrast, continuous high-cadence observations with Euclid are clearly powerful for constraining the microlensing parallax for short timescale events caused by low-mass lenses. We used the contours in Figure~\ref{fig:mag_te} together with joint $u_0$-$W149$ distributions for simulated WFIRST free-floating planet detections (Johnson et al. in prep.) to estimate that Euclid could measure parallaxes for 17\% of WFIRST's free-floating Jupiter-mass planets that it observes ($t_{\rm E}{\sim}1$~day), and 38\% of the $3$-Earth-mass planets it observes ($t_{\rm E}{\sim}0.1$~day). If the occurrence rate of free-floating planets follows the same distribution as bound microlensing planets, i.e.~\citep{Cassan2012},
\begin{equation}
    \frac{{\rm d}N}{{\rm d}M} =  \begin{cases} 0.24 (M/95 M_{\oplus})^{-0.73}~{\rm star}^{-1} & {\rm if}~M\ge 5 M_{\oplus}\\
    2~{\rm star}^{-1} & {\rm if}~M<5 M_{\oplus}\\
    \end{cases},
\end{equation}
then WFIRST can be expected to detect ${\sim}100$ free-floating planets per decade of mass between $2$ Earth masses and $2$ Jupiter masses, and ${\sim}50$ total free-floating planets with masses less than $2M_{\oplus}$. This number of detections corresponds to an expected detection rate of more than one free-floating planet detection every other day of observations~(Johnson et al. in prep). Of course, the actual occurrence rate of free-floating planets is unknown, but our adopted mass function is roughly commensurate with  planetary formation models seen in \citet{Veras2012} and \citet{Ma2016}.  A Euclid free-floating planet parallax measurement could therefore be expected once every ${\sim}6$~days of observations. In addition to this WFIRST is also expected to detect ${\sim}3$ bound planets every 24 hours~\citep{Penny2019}, so Euclid could provide a measurement of a bound or free-floating planet parallax once every 24 hours spent observing microlensing, if we assume a similar parallax detectability for bound planets. If Euclid were able to dedicate 20-days of observations to microlensing each time that WFIRST was also observing (60 days during its prime mission, and 60 days during an extended mission), it could measure parallaxes for ${\sim}20$ free-floating planets and ${\sim}60$ bound planets.

%{\color{red} Discussion of high-cadence (30-min), free-floating planet parallax detections and detection rates here.}

\begin{figure}[ht!]
\plotone{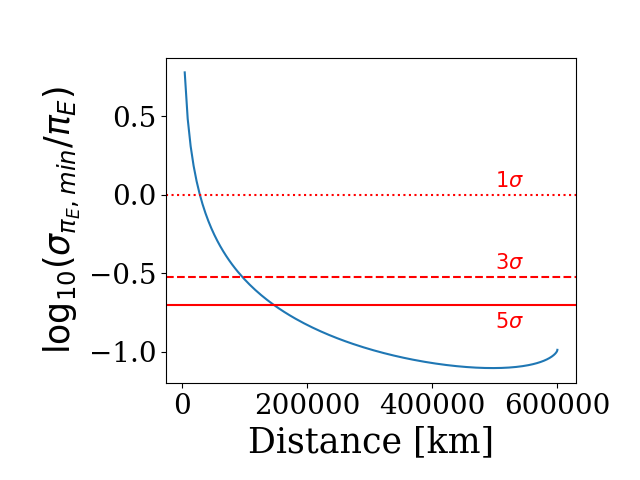}
\caption{Parallax measurement precision as a function of distance between the WFIRST and Euclid. Small-dashed, long-dashed and plain horizontal lines show 1,2 and 3 $\sigma$ detection. For this simulation, $u_0=0.1$, $t_E=3$~d, and the baseline photometric precision is $\sigma_{WFIRST}=0.01$ mag and $\sigma_{Euclid}=0.005$ mag, corresponding to a $W146=20$ mag source. Euclid cadence is set to 1 day. \label{fig:eucl_wfirst_phi}}
\end{figure}

Finally, we consider the impact of the relative separation of Euclid and WFIRST in their L2 orbits by varying the orbital phase difference between the two spacecraft that otherwise occupy the same orbit. If the projected separation of the two spacecraft is too small, then Euclid would only provide a small increase to the signal to noise ratio of the annual and L2-orbital parallax, without contributing a much stronger constraint via satellite parallax. Figure~\ref{fig:eucl_wfirst_phi} shows the parallax uncertainty as a function of the distance between spacecrafts for a microlensing event with $u_0 = 0.1$, $t_E = 3$~d and Euclid cadence of $1$~d. It is clear that even a small distance (e.g., $d>100,000$ km) between the telescopes provides enough constraint for the detection of the microlensing parallax. The required separation shrinks as the lens mass decreases.

\section{Conclusions} \label{Discussion and Conclusions}

In this letter we have considered the ability of the Euclid spacecraft to provide microlensing parallax measurements by observing contemporaneously with the WFIRST microlensing survey. Such measurements will be possible provided the projected separation between Euclid and WFIRST is larger than $100,000$~km. We find that high-cadence observations by Euclid for sustained periods of time as short as one day (i.e., longer than the timescale of a free-floating planet event) enable the measurement of parallaxes for large numbers of WFIRST's free-floating planets with short timescales. A significant subset of these would also have finite source measurements of their angular Einstein radii, allowing accurate mass measurements that would be impossible other than for a much smaller number of free-floating planet events that could also be observed from the ground. For each day of simultaneous high-cadence Euclid observations we would expect $0.17$ free-floating planet parallax measurements. With current launch schedules, Euclid and WFIRST could observe simultaneously for a total of 60 days during Euclid's primary mission (less than three percent of the total primary mission), and another 60 days during an extended Euclid mission, which would enable parallax measurements for $20$ free-floating planets and many more bound planets. For free-floating planets there is no other way to measure their masses.

Due to Euclid's earlier launch date, WFIRST is likely to begin observing toward the end of Euclid's main mission, when the likelihood of schedule holes is high~\citep{GomezAlvarez2018}. Once Euclid's main mission is over, it may be possible to significantly relax Euclid's solar aspect angle constraints which are designed to maintain the spacecraft's thermal stability. This could significantly increase the length of time Euclid could continuously observe microlensing for, and increase the number of free-floating planets for which it can measure parallaxes and masses.

We found that deep, dithered, low-cadence observations by Euclid, to be taken in holes in Euclid's regular observing schedule could provide parallaxes for a modest number of microlensing events with timescales. While these observations would go deeper than high-cadence observations in co-added exposures, they would not allow measurement of parallaxes for short-lived and rapidly-evolving free-floating planet events. In contrast, while having a higher ratio of overhead to observations, high-cadence observations would be randomly dithered and could be co-added for depth, while retaining the time-sampling necessary for detecting short-lived free-floating planet events that last only a few hours. We therefore recommend that any microlensing observations taken during gaps in Euclid's schedule be taken in a high-cadence mode, even if the gap is only a few hours long. 

We have only considered the ability of Euclid to provide free-floating planet parallaxes, but the same principles also allow the measurement of parallaxes for bound planets at similar rates. Additionally, high-cadence two-filter Euclid observations would allow the discrimination of potential exoplanet false-positives caused by microlensed binary sources~\citep[e.g.][]{Gaudi1998} and for tens of thousands of transiting planets that can be discovered in both Euclid and WFIRST microlensing observations~\citep{McDonald2014,Montet2017}; in both scenarios false positives have chromatic lightcurves, while planets have achromatic lightcurves. Simultaneous Euclid observations could also be used to measure the colors of faint Kuiper belt objects and confirm asteroseismic measurements of bulge giants~\citep{Gould2016}. Microlensing surveys also have the potential to detect exo-moons \citep{Han2008,Liebig2010}. For these objects, complimentary observations from Euclid to WFIRST would be very valuable to break model degeneracies seen in the past \citep{Bennett2014,Hwang2018}, in addition to the mass measurement. Finally, Euclid's VIS observations would significantly improve WFIRST's host star mass measurements via direct-detection of the lens by providing flux measurements in a different bandpass, and measurements of color-dependent centroid shifts~\citep[e.g.,][]{Bennett2007,Bhattacharya2018}. Also, the clear distinction between a free-floating planet and a wide binary (i.e., $\ge10$ AU) from the analysis of the lightcurve alone is challenging \citep{Mroz2019,Henderson2016b}. While one can expect the photometry from WFIRST and Euclid would be precise enough to reveal the subtle asymmetries in the lightcurves due to wide binaries, a detailed study of the blend flux and the blend proper motion in various bandpass in addition to the parallax effects are a clear benefit of joint Euclid observations. Each of these effects is beyond the scope of this letter, but deserve further study.

%This study only presents one benefit of simultaneous observations between the two spacecraft to constrain the microlensing parallax. More interesting aspects of joint observations still need to be study, such as the measurement of the color centroid shift of events \citep{Bhattacharya2018}, but this is out of the scope of this letter.

%\section{Acknowledgements}
The authors are grateful for the referee comments that significantly improve the manuscript. Authors thank J.-P. Beaulieu, E. Kerins, J. Rhodes, R. Street, and S. Johnson for fruitful discussions. EB gratefully acknowledges support from NASA grants NNX15AC97G and 18-XRP18\_2-0016. MP acknowledges the support of NASA grant NNG16PJ32C.

\bibliographystyle{aasjournal}  % needs package natbib
\bibliography{microlensing.bib}



\end{document}